\documentclass[fleqn,usenatbib]{mnras}

\usepackage[T1]{fontenc}
\usepackage{ae,aecompl}
\usepackage[utf8]{inputenc}
\UseRawInputEncoding %as the first line of your source file.
\pdfoutput=1 
\usepackage[normalem]{ulem}
\usepackage{graphicx}	% Including figure files
\usepackage{amsmath, mathtools}	% Advanced maths commands
\usepackage{amssymb}	% Extra maths symbols
\usepackage{longtable, multicol}
\usepackage{rotate}
\usepackage{lscape}
\usepackage{natbib}
\usepackage{latexsym,verbatim}
\title[Hydrocarbons: degradation chemistry]{Interstellar Hydrocarbons: Degradation Chemistry in Diffuse Clouds}

\author[Awad \& Viti]{
Zainab Awad$^{1}$\thanks{E-mail:zma@sci.cu.edu.eg}
and Serena Viti$^{2,3}$\\
$^{1}$Department of Astronomy, Space Science, and Meteorology, Faculty of Science, Cairo University, 
Giza 11326, Egypt\\
$^{2}$Leiden Observatory, Leiden University, PO Box 9513, 2300 RA Leiden, The Netherlands\\
$^{3}$Department of Physics and Astronomy, University College London, London WC1E 6BT, UK
}

\date{Accepted XXX. Received YYY; in original form ZZZ}

\pubyear{2022}

\begin{document}
\label{firstpage}
\pagerange{\pageref{firstpage}--\pageref{lastpage}}
\maketitle

%%%%%% Abstract %%%%%%
\begin{abstract}
Observations of diffuse clouds showed that they contain a number of simple hydrocarbons (e.g. CH, C$_2$H, (l- and c-)C$_3$H$_2$, and C$_4$H) in abundances that may be difficult to understand on the basis of conventional gas-phase chemical models. Recent experimental results revealed that the photo-decomposition mechanism of hydrogenated amorphous carbon (HAC) and of solid hexane releases a range of hydrocarbons into the gas, containing up to 6 C-atoms for the case of HAC decomposition. These findings motivated us to introduce a new potential input to interstellar chemistry; the ``top-down" or degradation scheme, as opposed to the conventional ``build-up" or synthesis scheme. 
In this work, we demonstrate the feasibility of the top-down approach in diffuse clouds using gas-grain chemical models. 
In order to examine this scheme, we derived an expression to account for the formation of hydrocarbons when HACs are photo-decomposed after their 
injection from grain mantles. Then, we calculated the actual formation rate of these species by knowing their injected fraction (from experimental work) 
and the average rate of mantle carbon injection into the ISM (from observations). Our preliminary results are promising and reveal that the 
degradation scheme can be considered as an efficient mechanism for the formation of some simple hydrocarbons in diffuse clouds. However, an 
actual proof of the efficiency of this process and its rate constants would require comprehensive experimental determination.
\end{abstract}
%%%%% keywords %%%%%

%%% paper body text %%%%

\section{Introduction}
\label{sec:intro}
Diffuse atomic and molecular clouds are characterised by low gas densities (10 - 500 cm$^{-3}$), relatively high temperatures (30 - 100 K), and low visual extinctions (A$_v$ $\le$ 1 mag); hence they are transparent to interstellar ultraviolet (UV) photons that play a vital role in their chemistry (e.g. \citealt{snow06} and \citealt{tiel13} and references therein). Diffuse clouds can be considered as natural laboratories where astrochemists may examine the chemical evolution of the Universe; from the simplest atomic hydrogen (HI) to the most complex observed molecules such as Polycyclic Aromatic Hydrocarbons (PAHs; \citealt{shem12}) and fullerenes (e.g. C$_{60}^+$; \citealt{camp15}). 

The first observed molecules in the interstellar medium (ISM), indeed in diffuse gas, were detected about 80 years ago. These detections were assigned to CN and the two hydrocarbons; CH and then CH$^+$ \citep{swin37, mcke40}. See reviews by \citet{tiel13} and \citet{mcg18}, and references therein, for a historical background on the early observed species in the ISM and for a complete list of all identified species in the ISM, respectively. 

Since the first detection of hydrocarbons (hereafter HCs), many of them, including the radical CH \citep{shef08}, have been observed in diffuse regions and in diffuse gas in the line-of-sight to other environments.

Hydrocarbons are compounds made of only hydrogen and carbon atoms. They have two main categories; aliphatic and aromatic. Aliphatic hydrocarbons have a geometric structure as either linear chains branched or non-branched where carbon atoms are joined with single, double or triple chemical bonds or cyclic structure with only single bonded carbons (if the compound have 3 or more carbons). Aromatic hydrocarbons are those with a benzene, C$_6$H$_6$, ring as the unit of their structure. 

The cyclic forms of hydrocarbons are surprisingly spread in diffuse regions. \citet{cox88} recorded the first detection of the cyclic C$_3$H$_2$ (c-C$_3$H$_2$) in the diffuse clouds in the lines-of-sight towards galactic and extragalactic regions. They found that the species is a sensitive tracer for low density media and its abundance is comparable to that recorded for dark regions (10$^{-8}$ - 10$^{-9}$). \citet{cox88} concluded that c-C$_3$H$_2$ can survive in UV exposed regions (e.g. those with A$_v \le$ 1 mag). The linear form of this molecule (H$_2$CCC; hereafter l-C$_3$H$_2$) has been first identified in dark clouds \citep{cern91}, and then in the line-of-sight towards the massive star forming regions W51 e1/e2, W51D, and W49 with the ratio c-C$_3$H$_2$ / l-C$_3$H$_2$ ranges between 3 and 7 \citep{cern99}. More recently, \citet{luc2000} conducted a survey for diffuse and marginally translucent clouds that lie toward a sample of compact extragalactic millimetre (mm)-continuum sources and identified various hydrocarbons; namely C$_2$H, C$_3$H, C$_4$H, and c-C$_3$H$_2$. The survey results showed that C$_2$H and c-C$_3$H$_2$ are widely spread with relatively high abundances of 10$^{-8}$ and 10$^{-9}$, respectively, and a fixed ratio for C$_2$H / c-C$_3$H$_2$ of nearly 27. 
Similar results were obtained for the observed diffuse gas along all lines-of-sight toward the massive star forming regions (G34.3+0.1, G10.62-0.39, W51, W49N) and SgrA* \citep{ger11}. Afterwards, \citet{liz12, liz14} detected C$_2$H, C$_4$H, c-C$_3$H$_2$ and c-C$_3$H in diffuse clouds. All of these hydrocarbons have also been observed with comparable abundances in photon dominated regions (PDRs) and planetary nebulae (PNe) \citep{pety05, pety12, sch18}. 

Most of the hydrocarbons are ejecta from carbon-rich evolved stars in their asymptotic giant branch (AGB) stage. These compounds spread gradually into the ISM during the evolutionary stages of protoplanetary nebulae (PPN) and planetary nebulae (PN) to enrich the surrounding diffuse ISM with a wide spectrum of carbonaceous species \citep{alam03}. The absorption bands in diffuse ISM at 3.4, 6.85, and 7.25 $\mu$m are attributed to material with a significant amount of hydrogenated amorphous carbons; also known as HAC or a-C:H (e.g. \citealt{dul83} and \citealt{dart20} and references therein). The mechanism/scenario of the formation of hydrocarbons has been the concern of several studies whether observational, theoretical and/or experimental. The importance of the 3.4 $\mu$m band feature in tracing interstellar organic substances motivated \citet{men02} to perform experiments aimed to study the formation of C-H bonds in nano-sized carbon grains when exposed to a flux of atomic H. The results indicate that the formation of C-H bonds (i.e. the formation of hydrocarbons) may occur in a time scale 100 times shorter than the typical lifetime of a diffuse cloud (3 $\times$ 10$^7$ years) with a hydrogenation efficiency of 6\%. This value is higher than the required value, 1\%, to balance the destruction of these bonds by UV irradiation in diffuse clouds. Therefore, \citet{men02} concluded that hydrogenation of carbon grains can proceed under diffuse medium conditions and this process is very important to explain the existence of the 3.4 $\mu$m band feature in such clouds. 

\citet{pety05} used the IRAM Plateau de Bure Interferometer (PdBI) to observe several hydrocarbons, as well as CO and C$^{18}$O, in the horsehead PDR, and, using a one dimensional PDR model, they attempt to fit their observations. Unfortunately, none of their models was able to correctly reproduce the observed amounts of hydrocarbons. In order to improve their models, \citet{pety05} proposed three different scenarios but conclude that the most likely one is that the observed hydrocarbons result from the fragmentation of PAHs in the intense far-UV radiation due to the nearby star $\sigma$-Ori. 

From another perspective, \citet{ser08} modelled the evolution of carbon dust particles due to the passage of shock waves in warm interstellar and inter-cloud media. The authors assumed that HACs are the most abundant form of carbonaceous grain mantles and they should be considered the more realistic model of interstellar dust grains. Their results showed that HAC erosion in shocks can supply the ISM with hydrocarbon compounds three times more than the abundance supplied by graphite grains. \citet{ser08} also estimated that carbonaceous dust grains can survive in the ISM up to $\sim$ 2 $\times$ 10$^8$ years. 

More recently, experiments showed that the photo-dissociation of solid C$_6$H$_{14}$ and HAC mantles leads to the formation of simple hydrocarbons and small C-clusters \citep{dul15}. The authors were able to detect simple hydrocarbons such as C$_3$H$_2$, C$_6$H$_5$, C$_6$H$_6$ and other products on the form C$_n$H$_{2n-1}$ from the irradiation of HAC film by UV photons. 
The authors concluded that the detection of C$_3$H$_2$ among the products of the induced photo-decomposition of HACs may reflect a way for the formation of the widely spread c-C$_3$H$_2$ in diffuse ISM, and indicate the possibility of including the top-down (degradation) chemistry as an alternative process of the formation of such species in diffuse ISM. 
Another possibility to explain the existence of C$_2$H and C$_3$H$_2$ with relatively high abundances in diffuse gas in comparison to that found in planetary nebulae (factor of 1-10\%) is the ejection of these species from the PN to the surrounding diffuse gas \citep{sch18}. 

The failure of pure gas-phase models in reproducing and explaining the observed amounts of some HCs, e.g. C$_3$H$_2$, C$_4$H \citep{alat15, cuad15, guz15}, in diffuse clouds and the recent experimental results \citep{dul15} motivated us to model hydrocarbon chemistry in diffuse clouds by introducing the top-down mechanism to the classical gas-phase chemical models. This study aims also to investigate the sensitivity of hydrocarbon chemistry to the physical condition of the environment of diffuse clouds and will attempt to define the space parameter that leads to the best fit of observations. 

The layout of this paper is: Section \ref{sec:mod} explains the chemical model and the new approach we used to simulate the formation of hydrocarbons in diffuse interstellar environments. In Section \ref{sec:res} we display and discuss the key results of our models and the influence of the degradation chemistry (top-down mechanism) on the HC content in diffuse clouds. We also compare the model results with observations. Finally, Section \ref{sec:conc} summarises the main conclusions and remarks of this work.

%%%%%%%%%%%%%%%%%%%%%%%%%%%%%%%%%%%%%%%%%%%%%%%%%%%%%%%%%%%%%%%%%%%%%%%%%%%%%%%%%%%%%%%%%%%%%%%%%%%%%%%%%%%%%%%%%%%%
\section{Modelling Hydrocarbons}
\label{sec:mod}
We attempt to simulate the chemistry of hydrocarbons in a typical diffuse cloud (n$_H$ = 100 cm$^{-3}$ and T = 100 K) using a real-time-dependent gas-grain chemical model; {\sc uclchem}\footnote{The code website: https://uclchem.github.io/} \citep{hold17}. The code models a ``parcel'' of gas of homogeneous density, at a specific A$_v$ (determined by the user-defined density and the radius of the parcel). See \citet{hold17} for full details of the code. For this study we fix the interstellar radiation field (RF), (G$_0$), to  1 Habing\footnote{A habing = 1.2 $\times$ 10$^{-4}$ erg cm$^{-2}$ s$^{-1}$ sr$^{-1}$, which is equivalent to 10$^{8}$ photons cm$^{-2}$ s$^{-1}$.}. Hydrogen ions, H$^+$, are formed by the cosmic rays (CR) ionisation process that happens at the ISM standard ionisation rate; $\zeta_{\text {ISM}}$ = 1.3 $\times$ 10$^{-17}$ s$^{-1}$ \citep{wak05}. The gas-to-dust ratio is uniform so that one visual magnitude of extinction (A$_v$ = 1) corresponds to a total hydrogen column density of 1.6 $\times$ 10$^{21}$ cm$^{−2}$ \citep{dra11book}. 

At each time step, the model calculates the A$_v$ using the ``on the spot approximation":
$$A_v \sim n_H \times R ~/~ 1.6 \times 10^{21}$$ where $n_H$ is the volume density and $R$ the radius.

Table \ref{tab:initial} lists the initial physical and chemical conditions used for the reference model ({\it hereafter} RM) of this study. 
%%%%%%%%%%%%%%%%%%%%%%%%%%%%%%%%%%%%%%%%%%%%%%%%%%
%%%%%%%%%%%
% Table-1 %
%%%%%%%%%%%
\begin{table}
\caption{The initial elemental abundances and physical conditions used for the reference model (RM) of the present work.}
\label{tab:initial}
\centering
\begin{tabular}{lllll} \hline 
\multicolumn{2}{c}{\bf Initial abundances$^{\text a}$} &&\multicolumn{2}{c}{\bf Physical parameters$^{\text b}$} \\ \hline
Helium & 8.50 $\times$ 10$^{-2}$ && Density (cm$^{-3}$) & 100 \\
Carbon & 2.69 $\times$ 10$^{-4}$ && Temperature (K) & 100 \\
Oxygen & 4.90 $\times$ 10$^{-4}$ && Radius (pc) & 5.2 \\
Nitrogen & 6.76 $\times$ 10$^{-5}$ && A$_v$ (mag)& 1 \\ 
HAC$^{\dag}$ & 20\% n(C) &&  & \\
\hline 
\end{tabular}
\flushleft
$^{\dag}$ adapted for this study (see text). \\
References: (a) \citealt{asp09} , (b) \citealt{awad16}
\end{table}
%%%%%%%%%%%%%%%%%%%%%%%%%%%%%%%%%%%%%%%%%%%%%%%%%%%%%%%%%%%%%%%%%%%%%%%%%%%%%%%%%%%%%%%%%%%%%%%%%%%%%%%%%%%%%
\subsection{Degradation Chemistry}
\label{sec:top}
\citet{mulas13} showed that for lines-of-sight with interstellar extinction curves similar to the average interstellar extinction curve (ISEC), the abundance of carbon (in any neutral form; i.e. atomic or compound) locked into dust grains is about 20 part per Million; i.e. 2 $\times$ 10$^{-5}$. If we assume that in diffuse clouds along these lines-of-sight carbon atoms are removed from HAC mantles by shocks that occur at time intervals of one Million years, then the rate to produce the maximum abundance of carbon into the ISM via HACs injection ($R_{\text{inj}}$) can be expressed as follows
\begin{equation}
\label{eq:1}
R_{\text{inj}} = \frac{2 \times 10^{-5}}{3 \times 10^{13}}~ \text{n}_H = 6 \times 10^{-19} \text{n}_H  \quad \text{cm}^{-3} \text{s}^{-1}\\ 
\end{equation}
where n$_H$ is the total number of hydrogen in all forms in the cloud. 

In their recent time-of-flight (TOF) experiments, \citet{dul15} detected 20 different simple hydrocarbons, some are isomers, produced from the photo-decomposition of the evaporated solid Hexane (C$_6$H$_{14}$) and HACs from grain mantle analogues. In their Fig. 1, red arrows indicated the mass peaks corresponding to three of the resulted hydrocarbons and small carbon clusters. The authors expressed the presented HAC mass spectrum in units of equivalent number of carbon atoms. From this figure, we selected the strong signals and very crudely calculated the fraction, f$_X$, by which their assigned species are injected into the gas-phase after the decomposition of the HACs. A summary of these calculated fractions for the 12 assigned molecules is given in Table \ref{tab:frac}. 

The experimental data of the injected species show variation in the strength of their signal which may imply that the efficiency of the injection of HACs vary among species with an unknown efficiency factor; say `$ef$' where 0$< ef <$1. This factor is treated in our model as a free parameter and its value is the one by which we obtain the best fit to observations. In light of the total amount of carbons injected into the gas at the rate $R_{\text{inj}}$, expressed in Eq. \ref{eq:1}, the injection of any resulted molecule as a fraction f$_X$ of the original decomposed HAC will occur at an injection rate $R_{\text{inj}}$(X) given by the following equation
\begin {equation}
\begin{split}
\label{eq:2}
&
R_{\text{inj}}(X) = R_{\text{inj}} \times ef \times \text{f}_X \\
&
\qquad\qquad
= 6 \times 10^{-19} \text{n}_H  \times ef \times \text{f}_X  \quad\quad  \text{cm}^{-3} \text{s}^{-1}
\end{split}
\end{equation}

We included the degradation chemistry in our chemical network by introducing a set of 12 chemical reactions each occurring at the rate $R_{\text{inj}}$(X) derived from the TOF experimental data. These reactions are generally on the form
\begin{equation}
\label{eq:3}
\text{HAC} ~ \xrightarrow[R_{\text{inj}}(X)]{\text {inj}}~ \text{X}
\end{equation}
where the molecules X are those listed in Table \ref{tab:frac}. This route of formation is, then, added to the formation pathway of the species if it has any additional route in the gas-phase chemical network; see \S~\ref{gas} below.
%%%%%%%%%%%%%%%%%%%%%%%%%%%%%%%%%%%%%%%%%%%%%%%%%%%%%%%%%%%%%%
% note to the editor: this table has to be here after Eq. (1)%
%%%%%%%%%%%%%%%%%%%%%%%%%%%%%%%%%%%%%%%%%%%%%%%%%%%%%%%%%%%%%%
%%%%%%%%%%%
% Table-2 %
%%%%%%%%%%%
\begin{table}
\caption{The computed fraction, f$_X$, of HCs injected in the gas via the sublimation of HACs as assigned in Fig. 1 in \citet{dul15}.}
\label{tab:frac}
\centering
\begin{tabular}{lllll} \hline 
{\bf Molecule} & {\bf f$_X$} && {\bf Molecule} & {\bf f$_X$} \\ \hline
C     & 0.12 && C$_3$H$_7$  & 0.12 \\
CH$_4$   & 0.09 && C$_4$    & 0.06 \\
C$_2$H$_5$  & 0.06 && C$_5$H$_{11}$ & 0.03 \\
C$_2$H$_6^{\dag}$  & 0.09 && C$_6$H$_4$  & 0.03 \\
C$_3$    & 0.12 && C$_6$H$_5$  & 0.03 \\
C$_3$H$_4$  & 0.21 && C$_6$H$_6$  & 0.03 \\ \hline 
\end{tabular}
\flushleft
$^{\dag}$C$_2$H$_6$ is the shorthand of CH$_3$CH$_3$.
\end{table}
%%%%%%%%%%%%%%%%%%%%%%%%%%%%%%%%%%%%%%%%%%%%%%%%%%%%%%%%%%%%%%%%%%
\subsection{Gas-Phase Chemistry}
\label{gas}
The gas-phase network for our species list is taken from the UMIST Database for Astrochemistry (UDfA) database\footnote{UMIST website: http://udfa.ajmarkwick.net/} ratefile 2012 \citep{mce013} except for the species C$_3$H$_7$, C$_5$H$_{11}$, C$_6$H$_4$, C$_6$H$_5$ which were not included in this database. The gas-phase chemistry of both C$_3$H$_7$ and C$_3$H$^+_7$ included in our network is taken from the KIDA\footnote{KIDA website: http://kida.astrophy.u-bordeaux.fr/} database and reported in \citet{loi17}. The rest of the missing species (C$_5$H$_{11}$, C$_6$H$_4$, C$_6$H$_5$) have no reactions under the conditions of the ISM and are not included in the KIDA ratefile. For this reason, we produced a suitable simple chemical network for these species and for the new species that may further be formed in the medium. This network includes simple reactions such as dissociative recombination, ion-neutral, and photo-processes to account for the formation and destruction of the species following top-down routes. Table \ref{tab:chem} lists these reactions with their rate parameters from which the rate constants ($k$) are calculated self-consistently in the model using the UMIST formula (e.g. \citealt{woo07,mce013}) as follows:
$$
k = 
\begin{cases} 
\mbox{a) if two-body reaction}\\
\alpha ~ (T/300)^{\beta} ~ exp(-\gamma/T)~~ \mbox{cm$^3$ s$^{-1}$} \\ \\
\mbox{b) if CR ionisation reaction, for $\zeta = \zeta_{\text{ISM}}$}\\
\alpha \quad\quad \mbox{s$^{-1}$}\\ \\
\mbox{c) if photo-reactions}\\
\alpha ~ exp(-\gamma A_v) ~~ \mbox{s$^{-1}$} 
\end{cases} 
$$
where $\zeta$ is the CR ionisation rate, A$_v$ is the visual extinction and the constants $\alpha$, $\beta$, and $\gamma$ are the rate parameters. 

For the new chemistry in Table \ref{tab:chem}, and as an acceptable first approximation, we adopted the general Langevin rate constant. These rate constants are 10$^{-7}$ cm$^{-3}$ s$^{-1}$ and 10$^{-9}$ cm$^{-3}$ s$^{-1}$ for dissociative electron recombination (DR; AB$^+$ + e$^- \rightarrow$ A + B) and ion-neutral (IN; A$^+$ + B $\rightarrow$ C$^+$ +D) reactions, respectively \citep{oka03}. In the UMIST database, most of the photo-dissociation reactions (PH; AB + h$\nu \rightarrow$ A + B) have rate parameters $\alpha$ and $\gamma$ typically in the range 10$^{-10}$ -- 10$^{-12}$ s$^{-1}$ and 1 -- 3, respectively, and since they are temperature independent then $\beta$ = 0. For this exploratory calculation, it is reasonable to adopt the middle value of each range for the parameters; i.e. $\alpha \sim$ 10$^{-11}$ s$^{-1}$ and $\gamma$ = 2. We ran test models with different values of the parameters and found that the overall hydrocarbons chemistry is insensitive to changes in the parameters of this PH reaction set in Table \ref{tab:chem}. 

We note that the molecule C$_3$H$_4$ has two structural isomers CH$_3$CCH and CH$_2$CCH$_2$ with similar chemistries. We included both chemistries in our network, and the obtained results showed that the two species have almost identical chemical behaviours, with insignificant differences in their abundances, less than a factor of 2. Therefore, in this work, we displayed the results of one of the two species, CH$_3$CCH, chosen arbitrarily. In addition, we distinguished between cyclic (c-) and linear (l-) forms of the species such as the case of C$_3$H$_2$, where the linear form is written as H$_2$CCC in the UMIST masterfile. Reactions with the rate parameter $\alpha$ lower than 10$^{-13}$ s$^{-1}$ were excluded from the network to save computational time because these reactions will not have a significant contribution to the chemistry.

The full chemical network in the present work includes 123 species linked in a total of 1399 gas-phase reactions. The fractional abundances of any species X in the network is computed by solving a set of ordinary differential equation (ODEs), generally, on the form 
$$ \dfrac{d}{dt}(X) = \sum F_{rate}(X) - D_{rate}(X) $$ 
where F$_{rate}$(X) and D$_{rate}$(X) are the net rates of reactions of the formation and destruction of X, respectively. 
If the species X has an additional formation pathway through the degradation chemistry with a net rate of injection of R$_{\text {inj}}$(X), then this formula is updated to account for this extra formation route as follows 
$$ \dfrac{d}{dt}(X) = \sum F_{rate}(X)+ R_{inj}(X)- D_{rate}(X) $$

At any time step and after computing the fractional abundances of the species, the code (self-consistency) checks that the total abundances of all species containing a particular element do not exceed the initially defined elemental abundance of that particular element. For instance, in the case of C and HACs, we set an initial abundance of HACs to be 20\% of the total number of C initial abundance. At each time step, the code ensures that the total number of HACs does not exceed the 20\% and that the total abundance of C-containing species, including HACs, does not exceed the initial elemental abundance of C defined in the input and listed in Table \ref{tab:initial}.
%%%%%%%%%%%%%%%%%%%%%%%%%%%%%%%%%%%%%%%%%%%%%%%%%%%%%%%%%%%%%%%%%%%%%%%%%
% note to the editor: table 3 is related to section 2.2: gas-phase chem %
% Table-3: set of gas-phase chem                                        %
%%%%%%%%%%%%%%%%%%%%%%%%%%%%%%%%%%%%%%%%%%%%%%%%%%%%%%%%%%%%%%%%%%%%%%%%%
\begin{table*}
\caption{The set of top-down gas-phase reactions included in the model chemical network and their rate parameters to account for the formation and destruction of C$_5$H$_{11}$, C$_6$H$_4$, and C$_6$H$_5$, and their related species, that are missing in UMIST and KIDA astrochemical databases.} 
\label{tab:chem}
\centering
\begin{tabular}{llllllllll} \hline 
{\bf Reaction}&\multicolumn{2}{c}{\bf Reactants}&&\multicolumn{2}{c}{\bf Products}&&\multicolumn{3}{c}{\bf {Rate parameters}}\\ %\hline
{\bf Type$^a$}& {\bf Re1} & {\bf Re2} && {\bf Pr1} & {\bf Pr2} && {\bf $\alpha$} & {\bf $\beta$} & {\bf $\gamma$}\\ \hline
DR  & C$_3$H$_6^+$  &  E$^-$  &&  C$_3$H$_5$ & H && 1 $\times$ 10$^{-7}$ & 0 & 0\\ [0.8 ex]
    & C$_4$H$_6^+$  &  E$^-$  &&  C$_4$H$_5$ & H && 1 $\times$ 10$^{-7}$ & 0 & 0\\ [0.8 ex]
   & C$_5$H$_{10}^+$ &  E$^-$ &&  C$_5$H$_9$ & H &&  1 $\times$ 10$^{-7}$ & 0 & 0\\ [0.8 ex]

 & C$_5$H$_9^+$ & E$^-$ && C$_5$H$_8$ & H &&1 $\times$ 10$^{-7}$ & 0 & 0\\ [0.8 ex]
 & C$_5$H$_8^+$ & E$^-$ && C$_5$H$_7$ & H && 1 $\times$ 10$^{-7}$ & 0 & 0\\ [0.8 ex]
 & C$_5$H$_7^+$ & E$^-$ && C$_5$H$_6$ & H && 1 $\times$ 10$^{-7}$ & 0 & 0\\ [0.8 ex]
 & CH$_3$C$_4$H$^+$ & E$^-$ && C$_5$H$_3$ & H && 1 $\times$ 10$^{-7}$ & 0 & 0\\ [0.8 ex]
 & C$_5$H$_{12}^+$ & E$^-$ && C$_5$H$_{11}$ & H && 1 $\times$ 10$^{-7}$ & 0 & 0\\ [0.8 ex]
 & C$_6$H$_4^+$ & E$^-$ && C$_6$H$_3$ & H && 1 $\times$ 10$^{-7}$ & 0 & 0\\ [0.8 ex]
 & C$_6$H$_5^+$ & E$^-$ && C$_6$H$_4$ & H && 1 $\times$ 10$^{-7}$ & 0 & 0\\ [0.8 ex]
 & C$_6$H$_6^+$ & E$^-$ && C$_6$H$_5$ & H && 1 $\times$ 10$^{-7}$ & 0 & 0\\ [0.8 ex]
\hline 
IN & C$_3$H$_3^+$ & C$_2$H$_2$ &  & CH$_3$C$_4$H$^+$ & H &  & 1 $\times$ 10$^{-9}$ & 0 & 0\\ [0.8 ex]
 & C$_3$H$_7^+$ & CH$_3$CH$_3$ &  & C$_5$H$_{12}^+$ & H &  & 1 $\times$ 10$^{-9}$ & 0 & 0\\ [0.8 ex]
 & C$_5$H$_5$ & C$_3$H$_7^+$ &  & C$_5$H$_{12}^+$ & C$_3$ &  & 1 $\times$ 10$^{-9}$ & 0 & 0\\ [0.8 ex]
 & C$_2$H$_5^+$ & CH$_3$ &  & C$_6$H$_4^+$ & H &  & 1 $\times$ 10$^{-9}$ & 0 & 0\\ [0.8 ex]
 & C$_4$H$_3^+$ & C$_2$H$_3$ &  & C$_6$H$_5^+$ & H &  & 1 $\times$ 10$^{-9}$ & 0 & 0\\ [0.8 ex]
 & C$_2$H$_4^+$ & C$_4$H$_4$ &  & C$_6$H$_6^+$ & H$_2$ &  & 1 $\times$ 10$^{-9}$ & 0 & 0\\ [0.8 ex]
\hline 
PH$^b$ & C$_3$H$_7$ & PHOTON &  & CH$_3$CHCH$_2$ & H &  & 1 $\times$ 10$^{-11}$ & 0 & 2\\ [0.8 ex]
 & CH$_3$CHCH$_2$ & PHOTON &  & C$_3$H$_5$ & H &  & 1 $\times$ 10$^{-11}$ & 0 & 2\\ [0.8 ex]
 & C$_3$H$_5$ & PHOTON &  & CH$_3$CCH & H &  & 1 $\times$ 10$^{-11}$ & 0 & 2\\ [0.8 ex]
 & C$_5$H$_{11}$ & PHOTON &  & C$_5$H$_{10}$ & H &  & 1 $\times$ 10$^{-11}$ & 0 & 2\\ [0.8 ex]
 & C$_5$H$_{10}$ & PHOTON &  & C$_5$H$_9$ & H &  & 1 $\times$ 10$^{-11}$ & 0 & 2\\ [0.8 ex]
 & C$_5$H$_9$ & PHOTON &  & C$_5$H$_8$ & H &  & 1 $\times$ 10$^{-11}$ & 0 & 2\\ [0.8 ex]
 & C$_5$H$_8$ & PHOTON &  & C$_5$H$_7$ & H &  & 1 $\times$ 10$^{-11}$ & 0 & 2\\ [0.8 ex]
 & C$_5$H$_7$ & PHOTON &  & C$_5$H$_6$ & H &  & 1 $\times$ 10$^{-11}$ & 0 & 2\\ [0.8 ex]
 & C$_5$H$_6$ & PHOTON &  & C$_5$H$_5$ & H &  & 1 $\times$ 10$^{-11}$ & 0 & 2\\ [0.8 ex]
 & C$_5$H$_5$ & PHOTON &  & CH$_3$C$_4$H & H &  & 1 $\times$ 10$^{-11}$ & 0 & 2\\ [0.8 ex]
 & CH$_3$C$_4$H & PHOTON &  & C$_5$H$_3$ & H &  & 1 $\times$ 10$^{-11}$ & 0 & 2\\ [0.8 ex]
 & CH$_3$C$_4$H & PHOTON &  & C$_3$H$_2$ & C$_2$H$_2$ &  & 1 $\times$ 10$^{-11}$ & 0 & 2\\ [0.8 ex]
 & C$_5$H$_3$ & PHOTON &  & C$_5$H$_2$ & H &  & 1 $\times$ 10$^{-11}$ & 0 & 2\\ 
\hline
\end{tabular}
\flushleft
$^a$Reaction Type: DR: Dissociative Recombination, IN: Ion-Neutral, PH: Photo-process \citep{mce013} \\
$^b$ The rate constant of photo-reactions depends on the visual extinction (A$_v$) as $k$ = $\alpha ~ exp(-\gamma A_v)$ in s$^{-1}$.
\end{table*}
%%%%%%%%%%%%%%%%%%%%%%%%%%%%%%%%%%%%%%%%%%%%%%%%%%%%%%%%%%%%%%%%%%
\subsection{The model grid}
\label{sec:grid}
The reference model (RM) of this study is run under the standard physical conditions of a typical diffuse cloud with full injection rate ($ef$ = 1). Beside the RM, we ran a grid of 14 models, divided in four categories, in order to investigate the effect of (1) changing the environmental conditions namely; the density (models RO), the strength of the radiation field (models RF), and the CR ionisation rate (models Zeta), and (2) varying the injection rate of HACs into the gas (models EF) on the fractional abundances of the simple hydrocarbons produced in diffuse clouds. A list of the models can be found in Table \ref{tab:grid}. All models are run  utilizing the same chemical network and the fractional abundances are computed with respect to the total amount of H atoms in the medium in all forms. 
%%%%%%%%%%%%%%%%%%%%%%%%%%%%%%%%%%%%%%%%%%%%%%%%%%%%%%%%%%%%%%%%%%%%%
% note to the editor: table 4 is related to section 2.3: model grid %
% Table-4: model grid                                               %
%%%%%%%%%%%%%%%%%%%%%%%%%%%%%%%%%%%%%%%%%%%%%%%%%%%%%%%%%%%%%%%%%%%%%
\begin{table*}
\caption{The model grid performed to explore the effect of varying the physical conditions of the environment on the chemistry of the clouds. RM refers to the reference model of the study. }
\label{tab:grid}
\centering
\begin{tabular}{lccccc} \hline
{\bf Model} & \multicolumn{4}{c}{\bf The Variable Parameters} & {\bf Total No.}\\
{\bf Category$^{\dag}$} & {\bf Density} & {\bf RF (G$_0$)} & {\bf $\zeta_{\text {ISM}}^{\ddag}$} & {\bf $ef$}& {\bf of Models}\\ 
             & {\bf (cm$^{-3}$)} & {\bf (Habing)} & {\bf (s$^{-1}$)} &\\ \hline 
{\bf RM} & 100 & 1 & 1 & 1 & 1 \\ [0.8 ex]\hline
{\bf RO} & 10 -- 300 & 1 & 1 & 1 & 3 \\ [0.8 ex]%\hline
{\bf RF} & 100 & 0.1 -- 10 & 1 & 1 & 3 \\ [0.8 ex]
{\bf Zeta} & 100 & 1 & 0.1 -- 50 & 1 & 3 \\ [0.8 ex]
{\bf EF} & 100 & 1 & 1 & 10$^{-4}$ -- 1 & 5 \\ [0.8 ex] \hline \hline 
\end{tabular}
\flushleft
$^{\dag}$ Category: RO: the density, RF: the radiation field, Zeta: the CR ionisation rate, and EF: the injection efficiency\\
$^{\ddag} \zeta_{\text {ISM}}$ is the ISM standard CR ionisation rate (= 1.3 $\times$ 10$^{-17}$ s$^{-1}$).
\end{table*}
%%%%%%%%%%%%%%%%%%%%%%%%%%%%%%%%%%%%%%%%%%%%%%%%%%%%%%%%%%%%%%%%%%%%%%
\section{Results and Discussion}
\label{sec:res}
In this section we present and discuss the model results of the chemistry of a typical diffuse cloud applying the suggested degradation chemistry to the conventional pure gas-phase chemical models. The results are illustrated in Figs. \ref{fig:1} to \ref{fig:5} for two selected sets of hydrocarbons; the first ({\it hereafter} set-I) is the experimentally obtained set from HAC photo-decomposition \citep{dul15} and the second ({\it hereafter} set-II) includes  the observed species in diffuse clouds \citep{liz12}. 

\subsection{Impact of degradation chemistry}
\label{res1}
\begin{figure*}
%% trim left bottom right top
\includegraphics[trim= 1.0cm 0.5cm 1.0cm 0.5cm,clip=true,width=16cm]{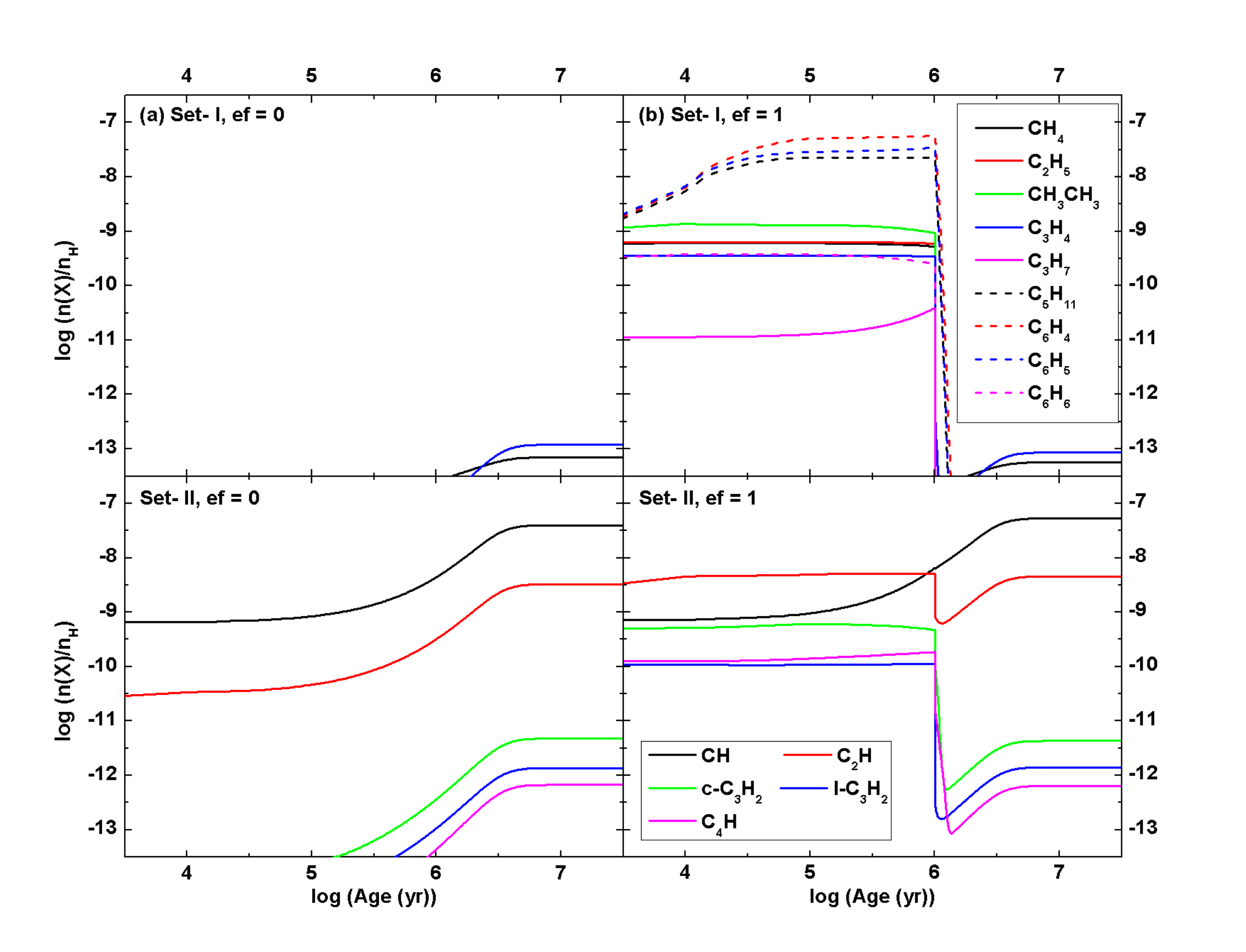} 
\caption {The influence of degradation chemistry on the chemical evolution of the gas in a typical diffuse cloud for two sets of species: in the top panel, species obtained in the results of \citet{dul15} experiments and in the bottom panel, the observed HCs by \citet{liz12,liz14}. The results are shown for two models: the classical gas-phase network without the top-down chemistry (column a) and the new RM model that combines the degradation chemistry with the classical gas-phase network (column b).}
\label{fig:1}
\end{figure*}

The impact of including the degradation chemistry to the classical gas-phase chemical models is illustrated in Fig. \ref{fig:1}. Column (a) of the figure shows the fractional abundances of the species using pure gas-phase models while column (b) is the classical gas-phase model with the inclusion of the degradation mechanism at maximum efficiency; i.e. $ef$ = 1. It is important to remember that the physical conditions of both models are those of the RM in Table \ref{tab:initial}: $\rho$ = 100 cm$^{-3}$, G$_0$ = 1 Habing and $\zeta$ = $\zeta_{\text {ISM}}$. The results are illustrated for both sets of HCs: set-I (the top panels) and set-II (the bottom panels), as quoted in the figure header. 
In general, and apart from CH in set-II, models with degradation chemistry show higher abundances of all HCs compared to the classical gas-phase models for both sets of species during early times (t $<$ 10$^6$ years). For both sets, the abundances in both models converges to similar values after 10$^6$ years. This time corresponds to the time when HACs reaches saturation and hence their injection becomes inefficient in the gas. By saturation we mean the time at which the amount of HACs removed from the grains reaches its maximum value and there is no renewal mechanism on grains.

During such early times, the inclusion of the top-down mechanism yields abundances of the order of 10$^{-9}$ -- 10$^{-10}$ for all the species in set-I except C$_5$H$_{11}$, C$_6$H$_4$ and C$_6$H$_5$ that have abundances of the order 10$^{-8}$ and C$_3$H$_7$ with a lower abundance of 10$^{-11}$. After that time, the abundances of all the species drop ($<$ 10$^{-13}$). The high abundances during the early stages of the evolution are due to the formation of the species via HACs injection into the medium. When the HACs abundances reach chemical saturation, t $\gtrsim$ 10$^6$ years, we obtain the sudden decline in the abundances of the injected species. Throughout the early stages of the evolution, formation and destruction processes are competing with comparable reaction rates which causes the evolutionary curves to remain relatively flat. After 10$^6$ years, atomic C reaches its maximum abundance and as a result, the role of HACs injection in the formation of the HCs in set-I becomes minimal. As a consequence, the formation rates drop leading to the dramatic decrease in the abundances of the HCs shown in Fig. \ref{fig:1}. CH$_4$ and C$_3$H$_4$ remain barely detectable in the gas with fractional abundances comparable to those of pure gas-phase models of $\sim$ 10$^{-13}$. 

We are aware that the species C$_5$H$_{11}$, C$_6$H$_4$ and C$_6$H$_5$ are those for which reliable reaction rates are in fact missing from the astrochemical databases and hence our results rely on our assumptions. 

A chemical analysis of the observed species CH, C$_2$H, l-C$_3$H$_2$, c-C$_3$H$_2$ and C$_4$H shows that CH is mainly formed by the reaction `H + CH$_2$' which has not been affected by the injection mechanism while the abundances of C$_2$H, l-C$_3$H$_2$, c-C$_3$H$_2$ and C$_4$H have been influenced, indirectly, by the injection chemistry. The analysis revealed that parent molecules of l-C$_3$H$_2$, c-C$_3$H$_2$ and C$_4$H are C$_3$H$_4$, C$_6$H$_3$ and C$_4$H$_3$, respectively. These latter species are either direct products of the HACs injection (e.g. C$_3$H$_4$) or they are daughters of other injected molecules (e.g. C$_6$H$_3$). This finding may explain why the molecular abundances of set-II drop at the same time as the species in set-I. For example, if we take the case of C$_6$H$_3$ we find that it is the tertiary product of the photo-decomposition of benzene, C$_6$H$_6$, which is a direct product of the photo-decomposition of mantle HACs \citep{dul15}. Therefore, when the production of C$_6$H$_6$ drops, as a consequence of the drop of HACs abundances, the abundance of C$_6$H$_3$ will drop; too. Other gas-phase formation routes for c-C$_3$H$_2$ will recover the abundance of the species to ultimately converge with that of pure gas-phase models ($ef$ = 0). 

However, the formation of C$_2$H in models with injection happens by many routes dominated by the photo-decomposition of C$_2$H$_2$ and C$_4$H. The analysis of the chemistry of both parents showed that they are daughters of species affected directly (CH$_4$ and C$_3$H$_4$) or indirectly (C$_4$H$_3$) by the decomposition of HACs. Thus, the abundance of C$_2$H drops when HACs injection drops because both formation routes via C$_2$H$_2$ and C$_4$H become minor. The recovery of the abundance of C$_2$H in the gas after 10$^6$ years is due to other gas-phase pathways such as dissociation of C$_6$H. Towards the end of the simulation, when the abundance of C in the medium increases and C reactions dominates, the formation of C$_2$H via C$_2$H$_2$ starts to become important again but with lower formation rates compared to earlier times (t $\le$ 10$^6$ yrs). This result may explain the convergence of both models illustrated in Fig. \ref{fig:1} at the late stages of the evolution. 

%%%%%%%%%%%%%%%%%%%%%%%%%%%%%%%%%%%%%%%%%%%%%%%%%%%%%%%%%%%%%%%%%%%%%%%%%
% note to the editor: Fig 1 is related to this section 3.1 %
% Fig. 1                                                   %
%%%%%%%%%%%%%%%%%%%%%%%%%%%%%%%%%%%%%%%%%%%%%%%%%%%%%%%%%%%%%%%%%%%%%%%%%
%%%%%%%%%%%%%%%%%%%%%%%%%%%%%%%%%
\subsection{Sensitivity to environmental parameters}
\label{res2}
%%%%%%%%%%%%%%%%%%%%%%%%%%%%%%%%%%%%%%%%%%%%%%%%%%%%%%%%%%%%%%%%%%%%%%%%%
% note to the editor: Fig 2 to 5 are related to this section 3.2 %
%%%%%%%%%%%%%%%%%%%%%%%%%%%%%%%%%%%%%%%%%%%%%%%%%%%%%%%%%%%%%%%%%%%%%%%%%
It is now established that variations in the environmental space parameters affect the chemistry of this environment and that the chemical evolution of a region may be used to constrain the physical parameters of these environments (e.g. \citealt{mil15, yama17}). In this section, we discuss the results of modelling HCs chemistry under various conditions of density, radiation field, CR ionisation rate, and injection rate that are denoted in Table \ref{tab:grid} by the model categories RO, RF, Zeta, and EF, respectively. 
In this study, an arbitrary factor of 8 was taken to account for the uncertainties associated with both observational values and many free parameters that go into the models. Hence, if the impact of varying a parameter in a given model is larger than 8 we consider its impact significant.

\begin{figure*}   
%% trim left bottom right top
\includegraphics[trim= 0.5cm 0.5cm 0.5cm 0.5cm,clip=true,width=18.5cm]{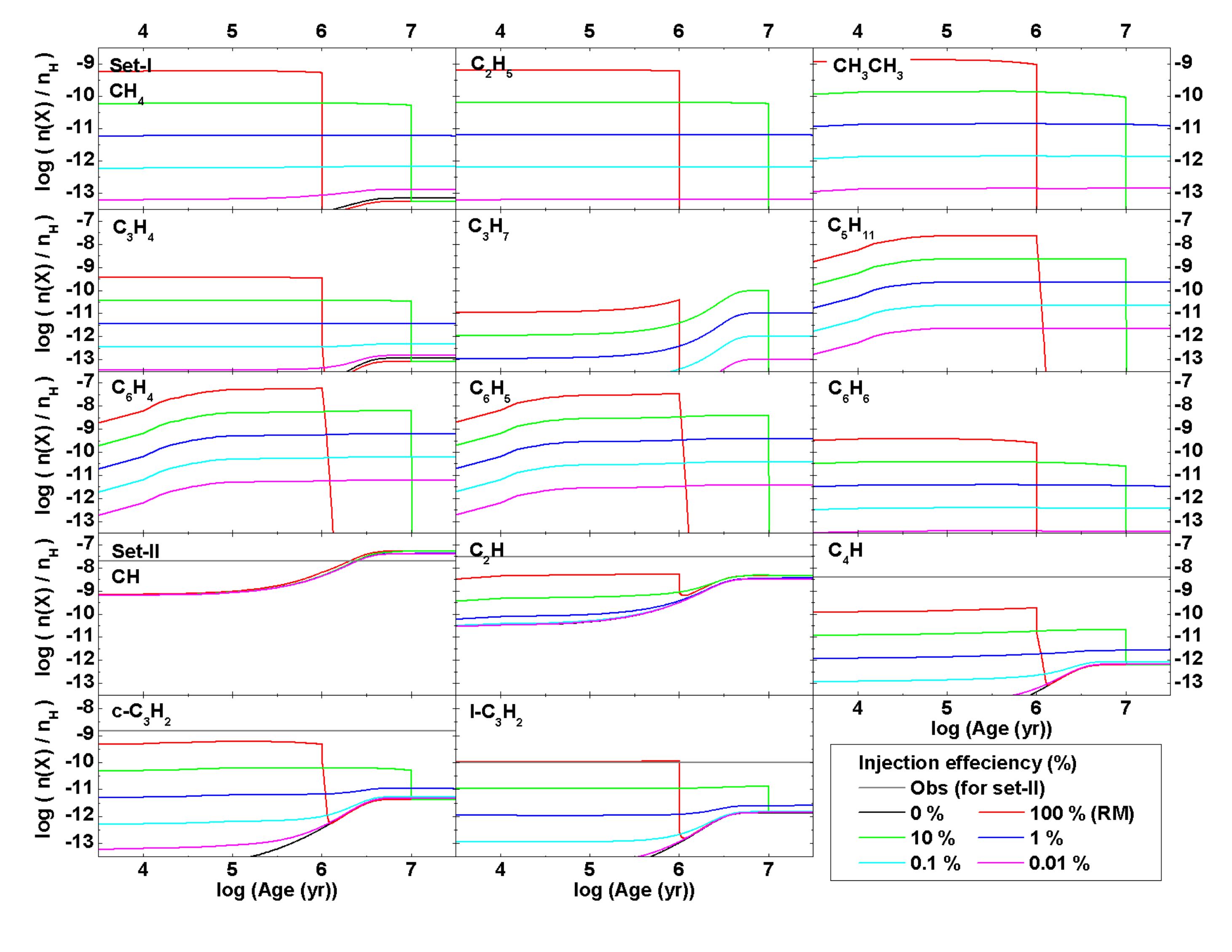}
\caption {The effect of varying the injection rate efficiency on the fractional abundances of the observed hydrocarbons in the diffuse gas. }
\label{fig:2}
\end{figure*}

\begin{figure*}   
%% trim left bottom right top
\includegraphics[trim= 0.5cm 0.5cm 0.5cm 0.5cm,clip=true,width=18.5cm]{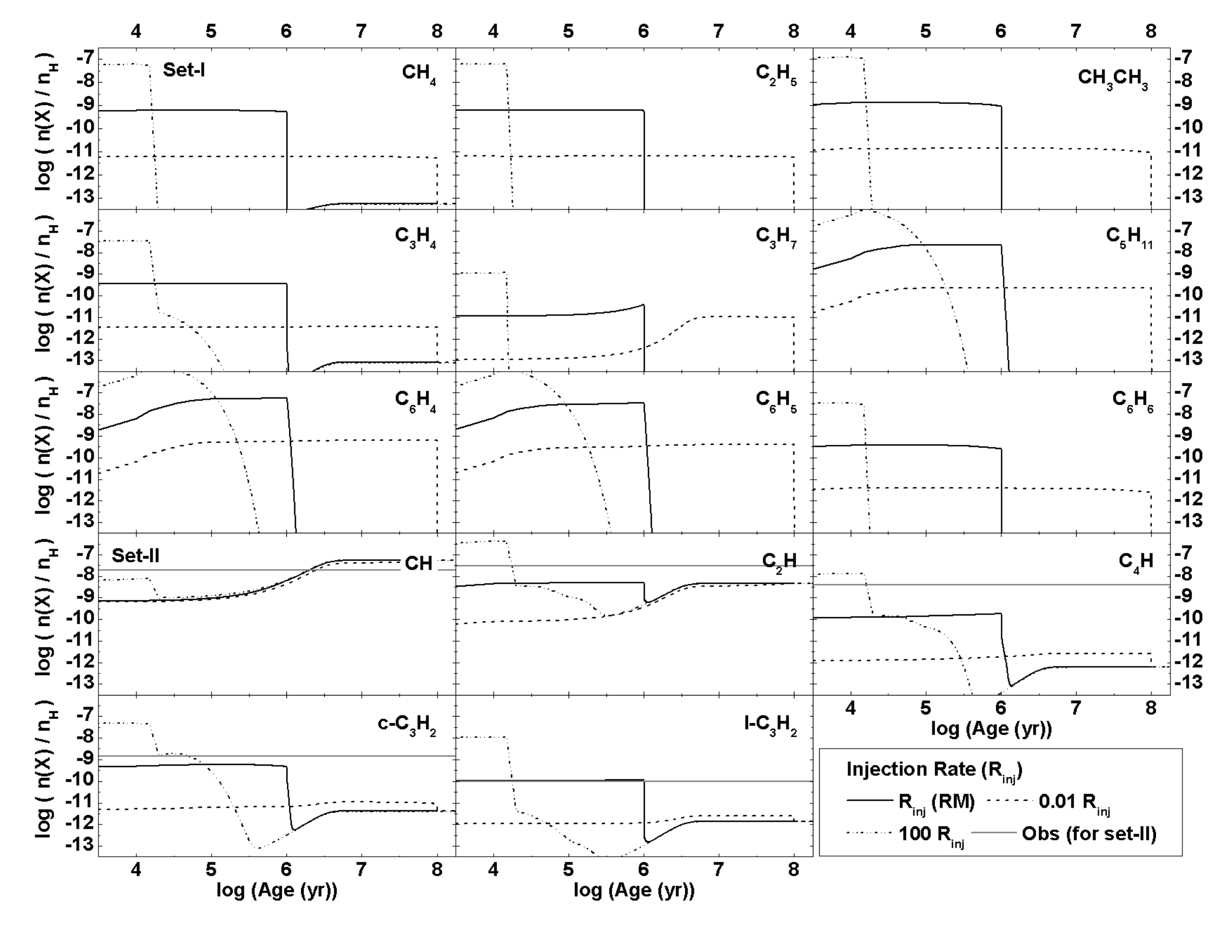}
\caption {The effect of varying the injection rate on the fractional abundances of the two sets of molecules in diffuse clouds. Each curve represents an injection rate; see the figure key.}
\label{fig:3}
\end{figure*}

It is interesting to investigate what fraction of HACs is needed to account for the formation of hydrocarbons in diffuse clouds. This motivated us to run chemical models with different injection rates, $R_{\text {inj}}$(X), category EF. According to Eq. \ref{eq:2}, variations in $R_{\text{inj}}$(X) are controlled by changes in the value of the parameter `$ef$'. 
Fig. \ref{fig:2} displays the results of the five models in the category EF plus the RM, for the two sets of species described earlier. All models are run under the same physical conditions of the RM (in Table \ref{tab:initial}), but at different injection rates, $R_{\text {inj}}$(X). Colour code is used to indicate different values of the parameter `$ef$'. The dark grey solid lines added to set-II panels represent the observed values of the species, taken from \citet{liz12}. As expected, increasing the injection rate, i.e. larger `$ef$' values, yields higher fractional abundances of species and show a better matching with observations for set-II molecules. An exception is CH where neither the evolutionary trend nor the calculated fractional abundances are affected by fluctuations in the injection rates. That is because the formation and destruction pathways of the species are independent of any products of the degradation chemistry as discussed in \S~\ref{res1}. 
%%%%%%%%%%%%%%%%%%%%%%%%%%%%%%%%%%%%%%%%%%%%%%
% Fig-2 %
%%%%%%%%%
%%%%%%%%%%%%%%%%%%%%%%%%%%%%%%%%%%%%%%%%%%%%%
The rate of removal of HACs from the grains, $R_{\text {inj}}$ in Eq. \ref{eq:1}, is computed assuming that the maximum amount of carbon is injected from grains by shocks every 10$^6$ years. Variations in the injection time scale is expected to impact the fractional abundances of the HAC products in the gas due to its impact on their production rate by injection, $R_{\text {inj}}$(X) in Eq. \ref{eq:2}. In order to test this idea, we performed two models that proceed under the same physical conditions of the RM but with different injection time scales. 
The first model (0.01 $R_{\text {inj}}$; indicated by dash lines in Fig. \ref{fig:3}) assumes that the maximum injection takes 100 Million years, thus the injection rate is 100 times slower than that of the RM ($R_{\text {inj}}$; solid black lines). The other model (100 $R_{\text {inj}}$; dash dot dot lines) proceeds with a maximum injection time that is 100 times shorter than the RM and therefore the injection rate is 100 times faster than the reference $R_{\text {inj}}$ in Eq. \ref{eq:1}. 

As expected, models with longer time scales (dash lines) always yield lower fractional abundances for all of the species in sets I and II compared to the RM (solid black lines). The reason for this is that with longer times of injection, the production rate of the injected molecules decreases by a factor of  100 compared to the RM. On the other hand, models with shorter time scales of injection (dash dot dot lines) lead to more material in  the gas compared to the RM model, in a shorter period and hence the lifetime of most of the species produced via 100 $R_{\text {inj}}$ models is shorter than that of the other two models. None of the two new models was able to reproduce the observed abundances of HCs, although the abundances of CH, C$_2$H, and C$_4$H in models with shorter injection time (i.e; 100 $R_{\text {inj}}$) are closer to observations (grey solid lines) during times $\le$ 10$^4$ years. For Set II molecules, the abundances and chemical evolution of CH are the least affected by changes in the injection time scales. All other molecules respond differently, but their obtained abundances in the three models tend to converge to values close to those of the RM at times later than 10$^6$ years; see Fig. \ref{fig:3}.
%%%%%%%%%%%%%%%%%%%%%%%%%%%%%%%%%%%%%%%%%%%%%%%%%%%%%%%%%%%%%%
% Fig- 3%
%%%%%%%%%
%%%%%%%%%%%%%%%%%%%%%%%%%%%%%%%%%%%%%%%%%%%%%%%%%%%%%%%%%%%%%%%

\begin{figure*}
%% trim left bottom right top
\includegraphics[trim= 0.8cm 1.0cm 0.5cm 0.5cm,clip=true,width=18.5cm]{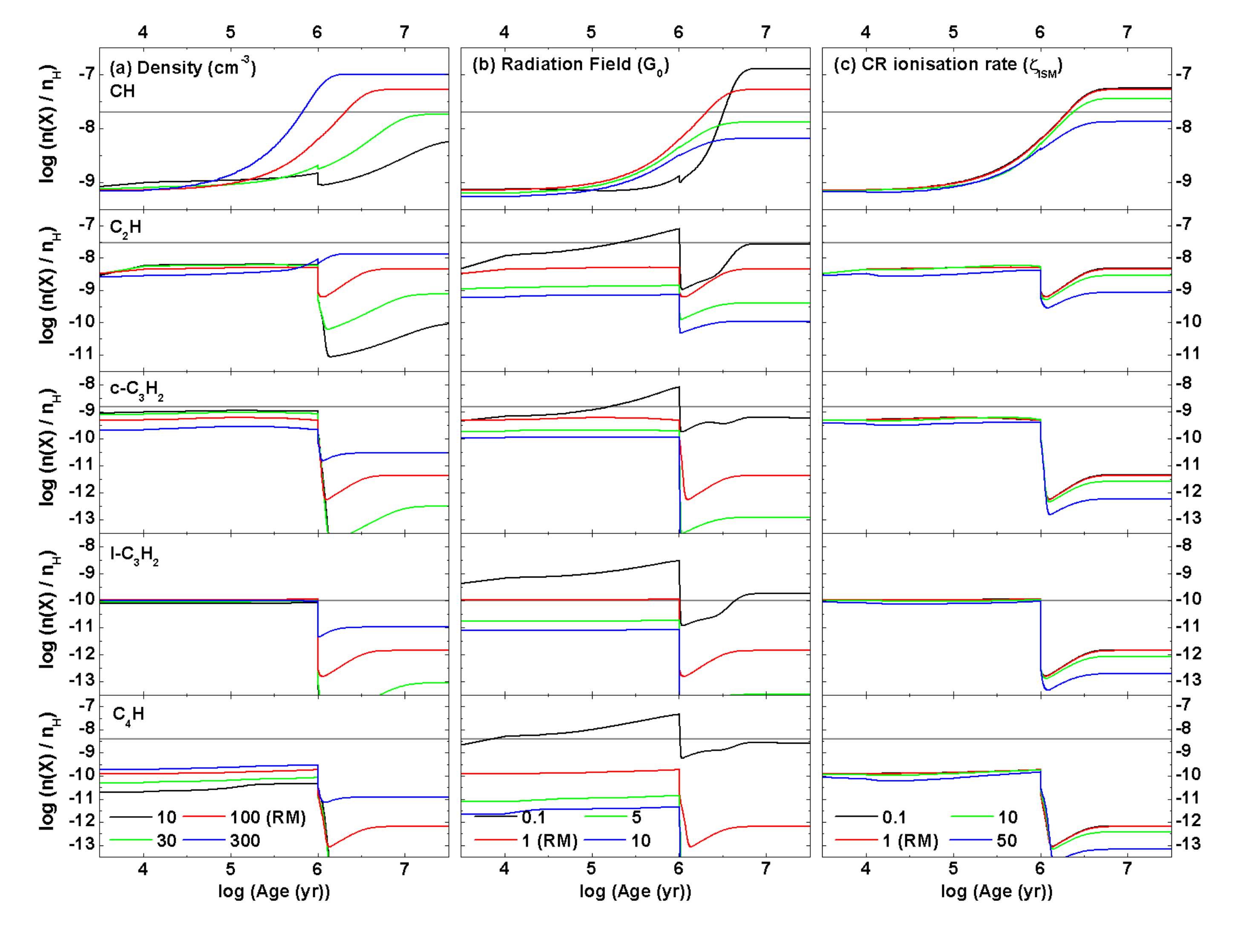}
\caption {The influence of changing the environment physical conditions on the fractional abundances of the observed set of species in diffuse gas. Different curve colours indicate different values for the physical parameters in each column: (a) The total gas number density in cm$^{-3}$, (b) the intensity of the radiation field in G$_0$, and (c) the CR ionisation rate in terms of the standard interstellar rate; $\zeta_{\text {ISM}}$, (see figure key). Dark grey lines mark the observations as taken from \citet{liz12}.}
\label{fig:4}
\end{figure*}

In order to qualitatively constrain the environmental conditions where HCs may exist, we modelled the chemistry of the observed hydrocarbons (set-II) under different environmental conditions. Fig. \ref{fig:4} shows the time evolution of the fractional abundances of these species for the different model categories in Table \ref{tab:grid}. Each column of the figure represents one model category labelled in its header. Different colours indicate different values of the same parameter of the model (see the figure key). Physical conditions that represent the reference model are denoted in the figure key as RM and represented by red curves. Dark grey lines mark the observed values as taken from Table 4 in \citet{liz12}. 

We find that all models show similarities in the evolutionary trends of the species with slightly different responses to changes in the space parameters. Although most of the species are insensitive to variations in the gas density (Column a) and CR ionisation rate (Column c) at times ($<$ 10$^6$ years), their steady state abundances (t $>$ 10$^6$ years) may differ by up to 2 orders of magnitude. In this study we are interested in the time interval when HACs feed the medium with hydrocarbons; i.e. 10$^4$ $<$ t (years) $\le$ 10$^6$. During this period, most of the abundances are enhanced in diffuse environments with lower UV photon intensity (0.1 G$_0$, black line). 

The calculated abundances of C$_2$H, c-C$_3$H$_2$, and C$_4$H in our RM are $\sim$ 6, 3.19 and 22 times less than those observed \citep{liz12}. In addition, \citet{liz18} showed that adding C-atom to C$_2$H ($\sim$ 4 $\times$ 10$^{-8}$) to form C$_3$H reduces the abundance of C$_2$H by about 100 times in all observed environments including diffuse regions; i.e. n(C$_3$H) $\sim$ of order 10$^{-10}$. Their observations of diffuse regions showed also that the ratio N(c-C$_3$H)/N(l-C$_3$H) is about 0.5 whereas N(C$_2$H) is 200 and 100 times higher than that of N(c-C$_3$H) and N(l-C$_3$H), respectively. The current network used in this work does not differentiate between c-C$_3$H and l-C$_3$H, but we can determine the total amount of C$_3$H in the gas. Our results show that, at 10$^6$ yrs, n(C$_3$H) is of order 10$^{-9}$ which is an order of magnitude higher than the approximate value observed. Adopting the ratio of \citet{liz18}, our calculations may indicate that n(c-C$_3$H) and n(l-C$_3$H) are 3 $\times$ 10$^{-10}$ and 6 $\times$ 10$^{-10}$, respectively. These values are comparable to those observed in the diffuse regions B0415 and B2200 (see Table 4 in \citealt{liz18}), although the ratios to C$_2$H are much less than the observed ratios. On the contrary, our computed values of orders 10$^{-9}$ and 10$^{-10}$ for C$_2$H and c-C$_3$H$_2$, respectively, are in good agreement with model calculations by \citet{guz15}, but for the ion C$_3$H$^+$ our value is 10 times less than their computed abundance at A$_v$ = 1 mag.

Of all the performed models, those with the following combination of physical parameters: n$_{H}$ = 100 cm$^{-3}$, $\zeta$ = $\zeta_{\text{ISM}}$, 0.1 G$_0$, and maximum injection of HACs yield a better match to observations although the abundances are 5 -- 30 times higher than the observed value \citep{liz12}; see Fig. \ref{fig:4} -- Column b. 
The chemical analysis of these species revealed that in the RM, most of those species are destructed in the gas-phase by UV photons and other less efficient pathways via ion-molecule reactions. In models with less intense radiation filed (= 0.1 G$_0$), the destruction of the species is dominated by ion-molecule reactions, in particular those involving C$^+$. These reactions may count for more than 70\% of the destruction pathways of a given species. The production of C$^+$ ions occurs, in general, via several photo-processes that may dominate the chemistry at low A$_v$. As a consequence, the reduction of the UV photon flux reduces the abundance of C$^+$ which, in turn, decreases the destruction rate of the species. Thus, an enhancement in the fractional abundances of HCs is obtained in diffuse clouds. 
%%%%%%%%%%%%%%%%%%%%%%%%%%%%%%%%%%%%%%%%%%%%%%%%%%%%%%%%%%%%%%%%%%%%%%
% Fig-4 %
%%%%%%%%%
%%%%%%%%%%%%%%%%%%%%%%%%%%%%%%%%%%%%%%%%%%%%%%%%%%%%%%%%%%%%%%%%%%%%%%%%
As already mentioned, there are many similarities in the evolutionary trend of all species, in particular C$_2$H, c-C$_3$H$_2$ and C$_4$H, species which have been extensively studied in PDRs \citep{pety05, mur20}. These similarities may indicate resemblance in the chemistry of diffuse clouds and PDRs regardless of their slightly different physical conditions. Moreover, \citet{alat15} studied experimentally the formation of hydrocarbons from the photolysis of HACs and then implemented their experimental results into PDR chemical models to study the effect of varying the UV radiation on the yield of the HCs. In their Fig. 4, we notice that their calculated abundances of C$_2$H, c-C$_3$H$_2$ and C$_4$H, at A$_v$ = 1, are comparable to our values within a factor of 10 at times $>$ 10$^6$ years. This factor becomes around 60 during the early stages of the evolution of c-C$_3$H$_2$ molecules. This supports our suggestion that there are similarities in the chemistries of diffuse clouds and PDRs that are easily penetrated with UV photons. These results also support our suggested new injection mechanism as an effective way to form small hydrocarbons in diffuse clouds. 

Moreover, \citet{pil12} studied experimentally the formation of molecules with double and triple carbon bonds from the irradiation of pure and mixed ices of c-C$_6$H$_{12}$ by highly charged and energetic ions that simulates cosmic rays particles. Their results showed that the maximum production of unsaturated hydrocarbons occur after (3-5) $\times$ 10$^6$ years for pure c-C$_6$H$_{12}$ ices and around 1 $\times$ 10$^6$ years for c-C$_6$H$_{12}$ mixed ices. The authors concluded that the CR bombardment of pure and mixed interstellar ice analogues can be an alternative pathway to the formation of hydrocarbons in astrophysical regions. Despite the difference in the mantle injection trigger between \citet{pil12} and this work, the results of our RM show that the maximum production of all studied hydrocarbons occurs at the time interval (1-2) $\times$ 10$^6$ years which corresponds to the time of maximum production of carbon in the gas via the HAC mantle injection. This agreement between the two production periods support our choice of the time interval of maximum injection and gives more support to the success of the degradation chemistry in producing interstellar hydrocarbons. 

\begin{figure*}   
%% trim left bottom right top
\includegraphics[trim= 1.0cm 0.5cm 1.0cm 0.5cm,clip=true,width=18.5cm]{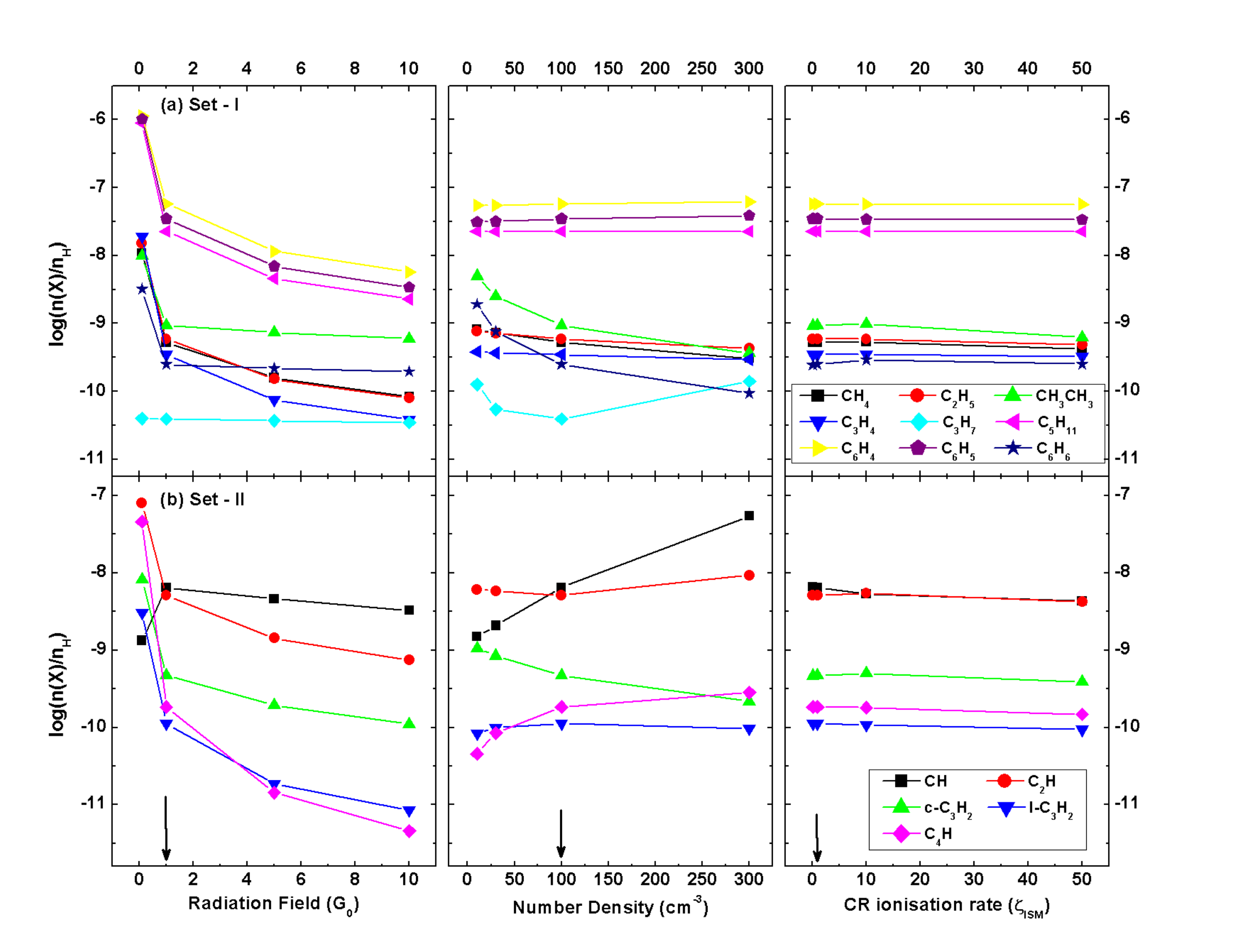}
\caption {The chemical evolution of the two sets of molecules as a function of environmental physical conditions (as indicated in each column header. The two sets of species are: the production of the photo-decomposition experiments by \citet{dul15} (top row) and the observed set in diffuse clouds (bottom row) by \citet{liz12}. Each curve denotes one molecule as labelled in the figure key.}
\label{fig:5}
\end{figure*}

To give an insight on the hydrocarbons chemistry and the relation with the surrounding environment, we demonstrate, in Fig. \ref{fig:5}, the chemical evolution of our selected species as a function of the different parameters at time 10$^6$ years which corresponds to the time of steady state of HACs (see above). From the figure, it is clear that hydrocarbons chemistry is most sensitive to variations in the radiation field (RF, left column). This result has to be expected because in diffuse clouds carbon hydrides (hydrocarbons) chemistry is initiated by the inefficient radiative association of C$^+$ and H$_2$ to form CH$_2^+$ which then undergoes various reactions with H, H$_2$ and electrons one of which will form the methyl group CH$_3^+$. The CH$_3^+$ recombines with electron to dissociate into simple hydrides, CH and CH$_2$ \citep{mil15}. From this discussion, we note that hydrocarbons chemistry is seeded by C$^+$ that is mainly formed by photo-reactions of atomic C and other carbonaceous species present in the ISM. Therefore, variations in the intensity of the radiation field affect the yield of C$^+$ and in turn influence the hydrocarbon chemistry. This finding and analysis of the hydrocarbons chemistry initiation in diffuse clouds may give an explanation on why hydrocarbons are insensitive to changes in the CR ionisation rate (right column). 

Unlike other species, CH showed an enhancement of its abundance at low values of radiation field (0.1 - 1 G$_0$). The parent molecule of CH is CH$_2$ which is formed through many photodissociation pathways. These routes are found to proceed with higher rates when the intensity of the UV photons is higher, yielding more CH$_2$ in the medium. This, in turns, enhances the abundance of CH in diffuse clouds that possess high radiation fields. In addition, we found that the formation rates of CH$_2$ are comparable across models with RF $>$ 1G$_0$, and therefore, the amounts of CH produced in these environments, at the same time step of the chemical evolution, are also comparable (see also Fig. \ref{fig:5}, middle column). The rest of the studied HCs are destroyed by C$^+$ whose abundance increases steadily with radiation field values up to 5 G$_0$, then it reaches a plateau. This finding may explain the observed decrease in the abundance of HCs with the increase of the RF to a value of 5G$_0$ before they reach an almost constant value. 

Maybe somewhat surprisingly, changes in the total number density of the gas appears to be selective (see Fig. \ref{fig:5}, middle column). While most species seem insensitive to changes in the number density, the abundances of CH, C$_4$H, CH$_3$CH$_3$, C$_3$H$_7$ and C$_6$H$_6$ show some sensitivity to it. The chemical analysis shows that the chemistry of these species is indeed linked. 
This may explain their evolutionary trends represented in Fig. \ref{fig:5}. We found that CH$_3$CH$_3$ is heavily destroyed via C$^+$ ions to form CH$_2$. The latter reacts with atomic H in the gas to produce CH. The formation rate of CH increases gradually with density  until it is about 2 orders of magnitude higher when the density increases from 10 to 300 cm$^{-3}$; this leads to the observed increase in the abundance of CH by $\sim$ 100 times its value in a 10 cm$^{-3}$ cloud. On the other hand and while the production rates of CH enhances with density increments, the destruction of CH$_3$CH$_3$ increases, too, causing its abundance to decrease in denser clouds. This may explain the trend observed for those species. In addition, increasing the gas density suppresses photo-reactions and decreases the rate of destruction of the species more than 10 times, such as the case for C$_4$H that shows an increase in its abundance in denser clouds. Similar scenarios are applicable for C$_3$H$_7$ and C$_6$H$_6$. 
%%%%%%%%%%%%%%%%%%%%%%%%%%%%%%%%%%%%%%%%%%%%%%
% Fig-5 %
%%%%%%%%%
%%%%%%%%%%%%%%%%%%%%%%%%%%%%%%%%%%%%%%%%%%%%%
\subsection{Comparison with observations}
\label{res3}
Model results showed that the evolutionary trend and the molecular abundance of the molecule CH remains unchanged under all modelling conditions, see Figs \ref{fig:2} and \ref{fig:5}. We note indeed that observations of CH in different diffuse molecular Galactic sight lines seem to imply a constant fractional abundance of $\sim$ 2.0 $\times$ 10$^{-8}$ \citep{shef08}. Modelling the formation of hydrocarbons by considering the top-down mechanism with full injection (i.e. RM model; $ef$ =1) show good agreement with observations for most of the species in set-II within a factor of 6. The results of C$_4$H are also reasonable since the observed value represents the upper limit of the molecule in diffuse clouds \citep{liz12}. The fractional abundances of set-II species match observations at times around 2 $\times$ 10$^6$ years.

We found that the calculated ratio C$_2$H/c-C$_3$H$_2$ is 11 at the time of maximum injection. This value is almost 2.5 times lower than the average of 27 seen in diffuse clouds \citep{luc2000} and in  PDRs \citep{pety05}. On the other hand, our value lies within the range obtained in star forming regions (11 to 53) although it is less than the quoted average of 28 \citep{ger11}. In addition, the computed l-C$_3$H$_2$/c-C$_3$H$_2$ in this work at 10$^6$ yrs is 0.23 which is 3 times the observed ratio in diffuse clouds \citep{liz12}, but matches the results of \citet{liz18}. However, this value is in agreement with the value derived from observations of diffuse gas towards the massive star forming regions W51 e1/e2 and W49N (0.2 - 0.3; \citealt{kul12}) in the low A$_v$ regime. Finally, the calculated abundance ratio C$_4$H / C$_2$H $\sim$ 0.04 is inline with the upper limit ratio, 0.14, observed by \citet{liz12}, and reported by \citet{liz18}.

Although the discussion of the results of this study is limited to the chemistry of hydrocarbons, we can also compare data for the injected small carbon clusters such as C$_2$ and C$_3$. The estimated fractional abundances for C$_2$ and C$_3$, at 10$^6$ years in the RM, are 3.9 $\times$ 10$^{-8}$ and 1.03 $\times$ 10$^{-9}$, respectively. These values are in good agreement with their corresponding average values, 3.0 $\times$ 10$^{-8}$ and 7.7 $\times$ 10$^{-10}$, observed towards the three objects HD 206267, HD 207198, and HD 210121 by \citet{oka03} and led to an abundance ratio of 40 which is, also, in good agreement with our calculated ratio of $\sim$ 38. 

Table \ref{tab:comp} summarises the calculated fractional abundances and ratios, in the RM at 10$^6$ years, for set-II species (CH, C$_2$H, l- and c-C$_3$H$_2$ and C$_4$H) and small carbon clusters (C$_2$ and C$_3$) when compared with observations in diffuse clouds. All observed fractional abundances with respect to H$_2$ are converted into abundances relative to the total H nucleons in the medium; where n(X)/n$_{\text H}$ = 0.5 n(X)/n$_{\text {H$_2$}}$ \citep{pety05}.
%%%%%%%%%%%%%%%%%%%%%%%%%%%%%%%%%%%%%%%%%%%%%%%%%%%%%%%%%%%%%%%%%%%%%
% note to the editor: table 5 is related to section 3.3: comparison %
% Table-5: model vs observations                                    %
%%%%%%%%%%%%%%%%%%%%%%%%%%%%%%%%%%%%%%%%%%%%%%%%%%%%%%%%%%%%%%%%%%%%%
\begin{table*}
\caption{A comparison between observations and the computed fractional abundances and ratios in the RM of this study, $x$(X), with respect to the total number of H in all form, n$_{\text H}$. In the table, a(b) is equivalent to a $\times$ 10$^b$.}
\label{tab:comp}
\begin{tabular}{llll} \hline 
            & {\bf Observations} & {\bf This work$^{\dag}$} & \\
{\bf Species}     & {\bf $x$(X)}  & {\bf $x$(X)} & {\bf Ref}\\ \hline 
%-------------------------------------------------------------------------------
{\bf CH}          &  20 (-9)      &  6.40 (-9) & \citealt{shef08}\\ [0.8 ex]
%-------------------------------------------------------------------------------
{\bf C$_2$H}      &  30 (-9)      & 5.10 (-9)  & \citealt{liz12}\\ [0.8 ex]
%                  &{\bf   4 (-8) }     &   & {\bf \citealt{liz18}}\\ [0.8 ex]
%-------------------------------------------------------------------------------
{\bf c-C$_3$H$_2$}&  1.5 (-9)    & 0.47 (-9) & \citealt{liz12}\\ [0.8 ex]
%-------------------------------------------------------------------------------
{\bf l-C$_3$H$_2$}&  0.1 (-9)     & 0.11 (-9) & \citealt{liz12}\\ [0.8 ex]
%-------------------------------------------------------------------------------
{\bf C$_4$H}  & 4.0 (-9)  & 0.18 (-9) & \citealt{liz12}$^{\ddag}$\\[0.8 ex]
%-------------------------------------------------------------------------------
{\bf C$_2$}   & 3.0 (-8) & 3.90 (-8) & \citealt{oka03}$^{\ddag\ddag}$\\ [0.8 ex]
%-------------------------------------------------------------------------------
{\bf C$_3$}  & 7.7 (-10) & 1.03 (-9) & \citealt{oka03}$^{\ddag\ddag}$\\ [0.8 ex] \hline
%-------------------------------------------------------------------------------
\multicolumn{4}{c}{\bf Hydrocarbons Ratios}\\ \hline
{\bf C$_2$H / c-C$_3$H$_2$}      &  27  & 11 &  \citealt{luc2000}\\ [0.8 ex]
%-------------------------------------------------------------------------------
{\bf l-C$_3$H$_2$ / c-C$_3$H$_2$}&  0.06  & 0.23 &  \citealt{liz12}\\ [0.8 ex]
%-------------------------------------------------------------------------------
{\bf C$_4$H / C$_2$H}& 0.14  & 0.04 &  \citealt{liz12}$^{\ddag}$\\ [0.8 ex]
%                    &{\bf  $<<$ 1 }& & {\bf \citealt{liz18}}\\ [0.8 ex]
%-------------------------------------------------------------------------------
{\bf C$_2$ / C$_3$}      &  40 & 38  &  \citealt{oka03}\\ \hline
\end{tabular}
\flushleft
{\bf $^{\dag}$} Model values are quoted at the time when HACs reaches saturation at 10$^6$ years. \\
{\bf $^{\ddag}$}The observed value is the upper limit. \\
{\bf $^{\ddag\ddag}$} The observed value is the average value in the three clouds {\bf HD 206267, HD 207198, and HD 210121}.
 \end{table*}
%%%%%%%%%%%%%%%%%%%%%%%%%%%%%%%%%%%%%%%%%%%%%%%%%%%%%%%%%%%%%%%%%%%%%%%%%%%%%%%%%%%%%%%%%%%%%%%%%%%%%%%%%%

\section{Conclusions}
\label{sec:conc}
Prompted by the experimental results of \citet{dul15}, we modelled the hydrocarbons chemistry in typical diffuse interstellar clouds by combining conventional gas-phase networks and a newly proposed degradation chemistry. We computed the rate of production of maximum number of carbons in the medium ($R_{\text{inj}}$) and the fraction (f$_X$) of hydrocarbon molecules produced from the photo-decomposition of HAC grains using the available experimental data. From these two quantities, we were able to express and determine the injection rate ($R_{\text{inj}}$(X)) of these hydrocarbons in the ISM and estimate the efficiency parameter of injection (ef). The influence of environmental variations on the hydrocarbons chemistry of the diffuse gas was also examined. 

Variations in the efficiency of the injection rate of the species leads to changes in both the calculated fractional abundances and the residence time of the molecules in the gas. Decreasing the injection rates, i.e. lower values of $ef$, reduces the computed abundances and increases the time interval of the survival of these species in the medium. The time after which the abundances of hydrocarbons steeply decrease corresponds to the time when HAC injection from grains becomes a minimum. Our models showed that we often need a maximum injection rate of carbons into the gas in order to reproduce observations. Our results also showed that the period when the hydrocarbon production reaches its maximum ($\sim$ 1-2 Million years) is in good agreement with the time obtained experimentally (3-5 Million years) from the CR bombardment of HAC films \citep{pil12}. 

In comparing our models with observations, we find that our RM model is capable of reproducing the abundances of most of the observed species in diffuse clouds within a factor of 6. Changes in the CR ionisation rates showed minor and/or insignificant variations in the abundances of all the set of studied species in the current work while variations in the number density appeared selective. However, the hydrocarbon chemistry was most influenced by fluctuations in the radiation field. The combination of physical parameters that better matches the observations are: gas total density of 100 cm$^{-3}$, temperature of 100 K, a standard CR ionisation rate and a low-intensity radiation field of 0.1G$_0$. This combination yields the observed abundances of all species \citep{liz12} within an order of magnitude accuracy. 

Our work highlights the importance of the inclusion of the degradation chemistry as a route of formation of hydrocarbons in diffuse clouds and emphasises the role of HAC mantles in enriching the ISM with such carbonaceous molecules. Hence, we may conclude that the degradation chemistry or the top-down mechanism is a new potential input to interstellar chemistry and chemical models. It is a promising scheme that enables chemical models of diffuse clouds to, successfully, produce comparable abundances of hydrocarbons to observations; unlike the case for pure gas-phase models. 
However, our present models are limited by uncertainties in the chemical networks of the species: C$_5$H$_{11}$, C$_6$H$_4$, C$_6$H$_5$. 
Further experimental work for a better understanding of the formation of hydrocarbons via top-down mechanism such as the estimation of decomposition and injection rates of hydrocarbons from HAC or other carbonaceous mantles, and the determination of the binding energies of the mantle fragments, is therefore desirable. 
%%%%%%%%%%%%%%%%%%%%%%%%%%%%%%%%%%%%%%%%%%%%%%%%%%
\section*{Acknowledgement}
This work arose from a suggestion made in conversation with David Williams, UCL. The authors are grateful for his fruitful discussion and valuable comments. 
%%%%%%%%%%%%%%%%%%%%%%%%%%%%%%%%%%%%%%%%%%%%%%%%%
\section*{Data availability}
The model outputs displayed in this article will be freely shared on request to the corresponding author.
%%%%%%%%%%%%%%%%%%%%%%%%%%%%%%%%%%%%%%%%%%%%%%%%%%
%\newpage
%%%%%%%%%%%%%%%%%
% References
%%%%%%%%%%%%%%%%%%
\bibliographystyle{mnras} 
%\bibliography{references}
%%%%%%%%%%%%%%%%%%

%%%%%%%%%%%%%%%%% APPENDICES %%%%%%%%%%%%%%%%%%%%%
%\newpage
%\section*{Appendix}
%\label{apn}
%%%%%%%%%%%%%%%%%%%%%%%%%%%%%%%%%%%%%%%%

% Don't change these lines
\bsp % typesetting comment
\label{lastpage}
\end{document}